\documentclass[preprint,showpacs]{revtex4}
\usepackage{graphicx}%
\usepackage{dcolumn}
\usepackage{amsmath}
\usepackage{latexsym}

\begin{document}
\title{Geometric Flows and Black Hole Entropy}  
\author{Joseph Samuel and Sutirtha Roy Chowdhury  }
\affiliation{Raman Research Institute, Bangalore-560 080}
\date{\today}
\begin{abstract}
Perelman has given a gradient formulation for the Ricci flow,
introducing an ``entropy function'' which increases monotonically
along the flow.
We pursue a thermodynamic analogy and apply Ricci flow ideas
to general relativity. We investigate whether
Perelman's entropy is related to (Bekenstein-Hawking)
geometric entropy as familiar from
black hole thermodynamics. 
From a study of the fixed points of the flow
we conclude that Perelman entropy is
not connected to geometric entropy. However, we notice that
there is a very similar flow which {\it  does} appear to be 
connected to geometric
entropy.
The new flow may find applications in black hole physics
suggesting for instance, new approaches to the Penrose inequality.

\end{abstract}

\maketitle



\section{Introduction}

The Ricci flow \cite{friedan,hamilton,perelman,cao} is an evolution equation 
for Riemannian geometries that tends to smooth and homogenise them. 
This flow is a (degenerate) parabolic differential equation, 
and is very similar to the heat equation.
This suggests a thermodynamic analogy
in which evolution of a geometry along the Ricci flow is like the approach 
of a physical system to thermodynamic equilibrium. 
In this paper, we pursue this analogy. We view the Ricci flow
from a thermodynamic perspective and  apply 
Ricci flow techniques to isolated gravitating systems
in general relativity (GR). Our motivation is to better understand 
physical quantities like energy and entropy in GR.

This paper is organised as follows. In section II, we
briefly describe Perelman's gradient formulation of the Ricci flow.
In section III, we motivate the application of these ideas to 
general relativity.
In section IV, we study fixed points of the 
Perelman flow and note that the Schwarzschild space is not a fixed point. 
In section V, we describe a slight modification of the 
Perelman flow. The fixed points 
of the new flow turn out to be initial data for the Schwarzschild black 
hole. Section VI generalises the discussion to include AdS.
Section VII is a concluding discussion.


\section{The Ricci Flow and Perelman's gradient formulation}
Let $(\Sigma,h_{ab})$ be a Riemannian manifold. 
($a,b$ run over 1,2,3. We restrict our discussion to three dimensional 
manifolds.)
Given an initial 
metric $h_{ab}$, the Ricci flow describes an evolution equation,
which evolves the metric according to its Ricci tensor. The evolution
parameter is $\tau$ and the family of metrics on $\Sigma$, $h_{ab}(\tau)$
satisfies the Ricci flow equation
\begin{equation}
\frac{\partial{h_{ab}}}{\partial \tau}=-2 R_{ab}
\label{ricci}
\end{equation}
In the neighborhood of a point $p\in \Sigma$, we can introduce 
a Riemann normal co-ordinate system and then the form of 
(\ref{ricci}) becomes parabolic ($\nabla^2$ is the Laplacian in
local co-ordinates)
\begin{equation}
\frac{\partial{h_{ab}}}{\partial \tau}= \nabla^2 h_{ab}
\label{heat}
\end{equation} 
and looks like a heat equation for the metric coefficients. However,
in a general coordinate system, the PDE (\ref{ricci}) is a {\it 
degenerate}
parabolic equation, because of its diffeomorphism invariance.

The Ricci flow first appeared in physics in the renormalisation
of $\sigma$ models\cite{friedan}. 
It was independently introduced in mathematics by Hamilton,
who used the flow to understand the topology of three manifolds.
Recently, Perelman has spectacularly used Ricci flow techniques to solve
the Poincare conjecture and it is widely
believed that these techniques also solve Thurston's
geometrisation conjecture.
These mathematical applications use the Ricci flow to address
questions of three dimensional topology. While Perelman's motivation
was topological, the techniques he introduced are tools of geometry 
and analysis. 
It seems reasonable to hope
that these tools will also find physical applications in
general relativity. The present paper is a preliminary step in this 
direction.
We suggest that the flow may be useful in understanding elusive
concepts like energy and entropy in general relativity. Both of 
these quantities are well understood in special relativistic
physics. In GR\cite{wald}, they take on a new global geometric interpretation, 
respectively the mass at infinity and the area of a horizon.

From a physical point of view, the most striking property of the Ricci
flow (which it shares with the heat equation) is its tendency to lose
memory of initial conditions. This is very similar to the
approach of an isolated physical system to thermal equilibrium.
For instance, a thermally insulated circular copper wire of circumference
$L$, which initially has a temperature distribution
$T(x), (T(0)=T(L),T(x)>0)$ evolves according to the heat
equation
\begin{equation}
\frac{\partial T}{\partial t}=\frac{d^2 T}{dx^2}
\label{heat2}
\end{equation}
(by choice of units, we set the diffusion constant to one)
and tends to a constant,
losing memory of the details of the intial distribution $T(x)$.
The final state is characterised by a single number, the uniform final
temperature $T_f$ and not an entire function $T(x)$. This ``information
loss'' is very reminiscent of entropy increase and indeed, one can 
view the heat equation in this light.
Consider the functional $S[T(x)]=\int_0^L dx a(T)$, where $a(T)=-T \log{T}$.
Using the heat equation (\ref{heat2}), integrating by parts and dropping
a divergence we see that $S[T(x)]$  
is monotonic  along the flow.
$\frac{dS}{dt}=\int dx a'(T){dT}/{dt}=
\int dx a'(T) {d^2T}/{dx^2}=-\int dx a''(T) ({dT}/{dx})^2\ge0$
The last inequality follows since $a(T)$ is a convex function $a''<0$.

The question we address here is:
can we in some sense regard the Ricci flow as representing the increase
of entropy and the approach of a gravitating system to its equilibrium
maximum entropy state. The question can be split into two parts.
1) Is there a functional of the metric which increases monotonically
along the Ricci flow? 
2) Does this functional physically represent geometric entropy?

One answer to the first question is provided by Zamolodchikov's C theorem,
proved in the context of two dimensional $\sigma$ models.
This theorem formalises the intuition that renormalisation is an 
irreversible process. The flow of coupling constants under 
renormalisation is such that a function $C(g_1,g_2...)$ of the coupling
constants of the theory is monotonically increasing along the flow.
In $\sigma$ models the role of the coupling constants is played by
the metric on the target space and the $\beta$ function is the Ricci
tensor\cite{friedan}. The C theorem guarantees that the RG
flow is irrotational and does not have limit cycles. However, the
Zamolodchikov C theorem appears to be a very specific result:
it applies only to two dimensional field theories and fails if the
target space is noncompact. This does not appear to be a general
answer to the question we raised.  
 
Another answer, which is more general and relevant to the
present context is provided by 
Perelman's gradient formulation of the Ricci flow. We briefly describe
this formulation. 
Consider the Ricci flow, supplemented by a diffeomorphism 
${\cal L}_{\xi}h_{ab}=D_a\xi_b+D_b \xi_a$
generated by a 
vector field $\xi^a$, which is itself a gradient $\xi_a=D_af$, where $f$
is a scalar function on $\Sigma$. 
\begin{equation}
\frac{dh_{ab}}{d\tau}=-2 R_{ab}+2 D_a D_b f
\label{perelmanflow}
\end{equation}
Perelman gave a gradient formulation for this flow by considering the
following ``entropy functional'' which depends on the pair
$(h_{ab},f)$, of tensor fields on $\Sigma$.
\begin{equation}
{\cal{F}_P}(h_{ab},f):=\int_\Sigma d^3x \sqrt{h}\, (\exp{f})\,[\,R+(Df)^2]  
\label{perelmanentropy}
\end{equation}
(In this paper $f$ is reversed in sign from Perelman's equations.)
There is a subtlety in the variation however: the degrees of freedom
in the metric can be split up into a conformal factor and a conformal
structure $[h_{ab}(x)]$, where the square brackets signify a
conformal class. In the variation of ${\cal F_P}$, the conformal structure
is freely varied, but the conformal factor is subject to the constraint
that the ``distorted volume''
$\sqrt{h}(\exp{f})$ is held fixed. 
Performing a variation
$\delta h_{ab}=\frac{dh_{ab}}{d\tau}\delta \tau$
and using the notation $\frac{dh^{ab}}{d\tau}=h^{ac} h^{bd}
\frac{dh_{cd}}{d\tau}$, we find
using a standard geometric identity for the variation of $R$,
\begin{equation}
\frac{dR}{d\tau}=-R_{ab} \frac{dh^{ab}}{d\tau}+D_aD_b\frac{dh^{ab}}{d\tau}
-D^2 (h_{ab}\frac{dh^{ab}}{d\tau})
\label{evolutionR}
\end{equation}
and after dropping some divergences.
\begin{equation}
\frac{d{\cal F_P}}{d\tau}=
\int_\Sigma d^3x \sqrt{h}\, (\exp{f})[-R_{ab}+D_aD_bf]
\frac{dh^{ab}}{d\tau}
\label{ratew}
\end{equation}

Thus $\frac{dh_{ab}}{d\tau}$ in (\ref{perelmanflow}) is twice the gradient 
of ${\cal F_P}$ subject to the constraint of perserving the distorted
volume.  It 
follows that
${\cal F_P}$ is non decreasing along the flow
\begin{equation}
\frac{d{\cal F_P}}{d\tau}\ge 0.
\label{nondecreasing}
\end{equation}
Hence the use of the word ``entropy'' for the Perelman functional.
The evolution equation for $f$ is not independent, but determined by the
constraint on the distorted volume. The reader is referred to
Topping \cite{topping} for a detailed exposition of Perelman's gradient
formulation.

In his paper on ``The entropy formula for the Ricci flow and its
geometric application'', Perelman remarks (section 5.3): ``The interplay 
between
statistical physics and (pseudo) Riemannian geometry occurs in the subject
of black hole thermodynamics developed by Hawking et al. Unfortunately,
this subject is beyond my understanding at the moment". Our objective here
is to pursue this remark and see if there is any connection between 
Perelman entropy and the geometric entropy familiar from 
black hole thermodynamics. This motivates our application of Ricci flow 
techniques to GR.


\section{Application to general relativity}
How does one apply Perelman's ideas to general relativity? The Perelman 
flow is defined in the space of three dimensional Riemannian metrics,
while GR deals with four dimensional Lorentzian ones. A natural approach
is to use spatial slices of four dimensional spacetimes  in order to define
the flow. The induced
metric on $\Sigma$ is a three dimensional Riemannian metric. In the
initial value formulation of GR, one uses the induced metric on $\Sigma$
and the extrinsic curvature of $\Sigma$ as initial data for the spacetime.
In this paper we restrict attention to time symmetric initial data.
The extrinsic curvature then vanishes and we need to deal only with
$(\Sigma,h_{ab})$ . Our idea is to apply the Ricci flow
to the initial data $h_{ab}$. For time symmetric initial data,
the dominant energy condition implies that the scalar curvature
$R$ is positive. The Ricci flow has the happy feature that it preserves 
the positivity of scalar curvature\cite{topping}. 
Thus, it may be used to understand general statements about initial data
that rely on such assumptions as positivity of curvature. Examples
of such statements are the positive energy theorem 
\cite{schoen,witten,woolgar} and the Penrose
inequality\cite{penrose,geroch,jang,huisken,bray,bray2}.
In all these examples Einstein's
equations do not
play any direct role: they are replaced by energy conditions, which
are geometric inequalities (for example, the positivity of scalar curvature).

In general relativity, energy and entropy are only well defined in
situations where the space has an asymptotic region. The total mass of an 
isolated system (the ADM mass) is defined using a fixed metric at 
infinity (either flat or AdS). 
The Bekenstein-Hawking entropy of black holes uses the notion 
of an horizon and ``escape to infinity''. It is therefore clear
that our $(\Sigma,h_{ab})$ must be a non-compact space, unlike
the compact situations considered by Perelman in the study of three
dimensional topology. Let us assume that $(\Sigma,h_{ab})$
has just one end at infinity and that $h_{ab}$ tends to a fixed
flat metric $\delta_{ab}$ at infinity \cite{wald}.  Even in ordinary
thermodynamics, entropy is only well defined in stationary
(or quasistationary) situations.
Since we assumed for simplicity that
we deal only with time symmetric situations, we can replace the word
`stationary' in the last sentence by static.
Correspondingly, in pursuing the
thermodynamic analogy we will restrict our attention to static,
four dimensional, Lorentzian, asymptotically flat space times,
subject to an Energy condition.

We now address the second question:
is the Perelman entropy related to geometric entropy, as
familiar from black hole physics? Perelman's entropy is 
non-decreasing along the flow.
However, the property of being monotonic along a flow is a fairly generic 
one and there may be many functionals which have this property. 
For instance,
in black hole physics, the area theorem asserts that the area of a horizon
can only increase with time. 
But then the same is true for the square of the horizon 
area.
In the example of the copper wire above, replacing $a(T)=-T\log{T}$
by any other convex function ($a''<0$) would also lead to a functional
which monotonically non-decreasing along the flow. Only the choice
$a(T)=\log{T}$ results in $S[T(x)]$ being proportional to the actual
thermodynamic entropy of the wire.
While all these ``entropy functionals''
differ, they share one property in common: they are all maximized
at the fixed point of the flow: the final state of uniform temperature.

Rather than look at the entropy function, it is more effective to
look at the fixed points of
the Perelman flow and ask if these correspond to maximum entropy in the
physical sense.
This motivates a study of
the fixed points of the Perelman flow. These
fixed points are referred to as `Ricci solitons'.

The stationary spacetime which maximises its geometric entropy for a
fixed energy\cite{sorkin},is the static, spherically symmetric
Schwarzschild black hole\cite{footnote1}. Black holes have enormously more 
entropy than
a star of the same mass (some $10^{20}$ times more for a solar mass). 
If Perelman's entropy were related to
geometric entropy, we would expect the Schwarzschild space
to be a fixed point of the Perelman flow. This expectation
can be tested by a direct
calculation which is presented in the next section.


\section{Ricci Solitons: Fixed Points of the Perelman flow:}
The end points of the Perelman flow are called Ricci solitons
and are characterised by a vanishing of the RHS of (\ref{perelmanflow}).
A Ricci soliton consists of a pair $(h_{ab},f)$ subject to the
equations:
\begin{equation}
R_{ab}=D_aD_b f
\label{fixedpoint}
\end{equation}
We wish to check if the Schwarzschild space
solves these equations for {\it some} choice of $f$. We can assume the
spherically symmetric form
\begin{equation}
ds^2=a(r)dr^2+r^2(d\theta^2+\sin{\theta}^2d\phi^2)
\label{spherical}
\end{equation}
for $h_{ab}$ and suppose that $f$ depends only on $r$. In order to 
preserve this gauge (\ref{spherical}) we will need to supplement the
RHS of (\ref{fixedpoint}) with an additional diffeomorphism
generated by a radial vector field
$U(r)\frac{\partial}{\partial r}$. Since this vector field is obviously
a gradient of a scalar function, we can absorb it in a redefinition 
of $f$. 

The independent fixed 
point (\ref{fixedpoint}) equations are $D^2f=R$ and $R_{\theta 
\theta}=(DDf)_{\theta 
\theta}$. These are two equations
for the pair $(a(r),f(r))$. We can use the second of these equations to 
eliminate $f(r)$ in favour of $a(r)$ and arrive at the autonomous 
differential
equation for $a(r)$.  
\begin{equation}
\frac{d^2 a}{d r^2}=\frac{3(a')^2}{2a}+\frac{2a^2-2a+ra'(1-a)}{r^2}
\label{ode}
\end{equation}
It is easily seen that (apart from the trivial case $M=0$) the
Schwarzschild space, which has $a(r)=(1-2M/r)^{-1}$
does not satisfy (\ref{ode}). Thus Schwarzschild
is {\it not} a fixed point of the flow {\it no matter what choice is made for
$f$}.
This shows that fixed points of the Perelman flow do not maximise
geometric entropy as familiar from black hole physics.
Flowing the Schwarzschild space using the Perelman flow would
lead to a {\it further} increase of Perelman's entropy.
We conclude that
Perelman's entropy function is not connected to Bekenstein-Hawking
entropy. However, as we argued earlier, many aspects of the Ricci flow
are strongly reminiscent of thermodynamics. Are there other similar flows
in the space of metrics which do describe the thermodynamics of geometry?
This question is addressed in the next section.


\section{The Modified Ricci flow and its fixed point}
Let us consider how one can modify the Ricci flow to make connections
with black hole entropy. We would like to have a geometric differential
equation, ({\it i.e} one constructed from tensor fields) of a parabolic
nature similar to the forward heat equation.  
The right hand side must contain no more than second derivatives 
of the basic fields $(h_{ab},f)$. Choosing the simplest possibility
consistent with these requirements,
we define the 
modified Ricci flow (MRF) by the equation
\begin{equation}
\frac{\partial h_{ab}}{\partial \tau}=-2fR_{ab}+2D_aD_b f
\label{newflow}
\end{equation}
This differs from Perelman's flow (\ref{perelmanflow})
only in the presence of $f$ in the first
term. $f$ can be interpreted as an effective space dependent diffusion
constant. It describes a forward type heat equation only in regions 
where $f$ is positive and bounded away from zero. In these regions we would
expect the flow to have the same behaviour as Perelman's flow, smoothing
out inhomogeneities and losing memory of initial conditions. Let us 
supplement the Ricci flow with a forward type evolution equation for
$f$:
\begin{equation}
\frac{\partial f}{\partial \tau}=D^2 f.
\label{modf}
\end{equation}
The pair $(h_{ab},f)$ evolved according to 
(\ref{newflow},\ref{modf}) constitute the modified Ricci flow 
\cite{footnote}.
Does the MRF
have any connection with geometric entropy?

As we mentioned before, the most efficient way to
investigate this question  is to study
the fixed points of the flow, characterised
by
\begin{equation}
fR_{ab}=D_aD_b f
\label{newfixed}
\end{equation}
where $f$ satisfies
\begin{equation}
D^2 f=0
\label{ffixed}
\end{equation}

It is easily seen that with 
\begin{equation}
f=(1-2M/r)^{1/2},
\label{schwarzf}
\end{equation}
the Schwarzschild exterior 
space
\begin{equation}
ds^2=(1-2M/r)^{-1}dr^2+ r^2(d\theta^2+\sin{\theta}^2d\phi^2)
\label{schwarzschild}
\end{equation}
($r>2M$) is a fixed point of the MRF.
This is the second main observation we make
in this paper.
The modified Ricci flow does seem to have
some connection with geometric entropy: its fixed points are
extrema of the Bekenstein-Hawking entropy. Note that the function
$f$ is positive in the Schwarzschild exterior and that it degenerates
to zero on the inner boundary $r=2M$. 
The form of $f$ motivates the identification of the pair
$(h_{ab},f)$ with the static spacetime
\begin{equation}
ds^2=-f(x)^2dt^2+h_{ab} dx^a dx^b
\label{static}
\end{equation}
This gives a Lorentzian spacetime
interpretation of $f$ as the lapse function or the redshift factor. 

\section{Generalisation to AdS}
For the compact spatial topologies studied by Perelman, 
there was room to modify the RHS of the 
Ricci flow by the addition
of a ``cosmological constant'' $\lambda h_{ab}$. This led to the
possibility of additional Ricci solitons, shrinking and expanding breathers.
In the last two sections, 
we required that our metric $h_{ab}$ tend to a {\it fixed flat} metric at 
infinity and so $\lambda$ had to vanish to preserve the asymptotic 
conditions. It is interesting to ask what happens if we replace
asymptotic flatness by asymptotically AdS. The arguments of the 
last two sections go through with appropriate slight
modifications. In the equations (\ref{perelmanflow},\ref{fixedpoint}),
one simply replaces the Ricci tensor
$R_{ab}$ by $R_{ab}-\lambda h_{ab}$, where $\lambda$ is a 
negative cosmological constant. Correspondingly, (\ref{ode}) acquires an
extra term $\lambda raa'$ on the right hand side. As before we see that
the AdS Schwarzschild space, which has $a(r)=(1-2M/r-\lambda r^2/3)^{-1}$
does not satisfy the soliton equation.

However, we modify the flow to read
\begin{equation}
\frac{\partial h_{ab}}{\partial \tau}=
-2f(R_{ab}-lambda h_{ab})+2D_aD_b f
\label{newflowlam}
\end{equation}
supplemented by an evolution equation for $f$:
\begin{equation}
\frac{\partial f}{\partial \tau}=D^2 f+\lambda f.
\label{modflam}
\end{equation}
Then we find that the AdS 
Schwarzschild is indeed a Ricci Soliton of the modified flow
with $f$ given by
$f=(1-2M/r-\lambda/3 r^2)^{1/2}$.
Thus our main conclusions generalise easily to include AdS space.


\section{Conclusion}
In his paper\cite{perelman}, Perelman remarks that the interplay 
between statistical physics and (Lorentzian) geometry occurs 
in the subject of black hole thermodynamics. He appears to suggest
possible connections between his entropy functional and black 
hole thermodynamics. 
Perelman was working with compact Riemannian manifolds. It
seems clear that for a discussion  of geometric entropy we need to
have non-compact manifolds, so that we can define a region of no
escape. The main result of this paper is the observation that
the Perelman entropy is {\it not}
the geometric entropy, at least if one uses the natural identications
we have made between initial data and spacetimes. 

Another observation we make is that a small modification
of Pereman's flow does seem to relate to geometric entropy. 
It would seem important to understand the properties of the 
new flow and possibly relate it to the Perelman flow. For instance,
What can one say about the long term behaviour of both flows
starting with an arbitrary initial positive curvature metric
in the asymptotically flat context. How does one choose the initial
$f$? Does the MRF admit a gradient formulation?
We are presently studying both flows in the
linear regime in the neighborhood of their Ricci solitons.

Several papers have appeared \cite{oliynyk,oliynyk2,headrick,bakas} in the physics literature 
about physical applications of the Ricci flow. Many of them are motivated
by the renormalisation group, string theory or conformal field theory.
Our motivation is from the viewpoint of classical general relativity.
In GR, there are a number of results of a purely
differential geometric character which also have a strong thermodynamic
flavour. Examples abound in black hole physics, but let us mention
two: the Penrose inequality and the Hawking area theorem. 
These results do not appeal to Einstein's equations but only to 
energy conditions. The Ricci flow techniques of Perelman do seem
to be a natural tool to attack such problems.

 


\begin{thebibliography}{1}
\bibitem{footnote1} We are suppressing a subtlety here. Quantum
effects lead to black hole evaporation and a further increase
of entropy. This paper is concerned only with classical GR.
\bibitem{footnote}
The term ``modified Ricci flow'' is sometimes used in the mathematical
literature to describe Ricci flow with surgery. This is not to 
be confused with the flow described here.
\end{thebibliography}
\end{document}